\documentclass[12pt,preprint]{aastex}
\usepackage{emulateapj5,epsfig}
\newcommand{\LCDM}{$\Lambda$CDM}
\newcommand{\WMAP}{{\it WMAP}}
\newcommand{\SDSS}{{\it SDSS}}
\newcommand{\msun}{{M_{\odot}}}
\newcommand{\hmsun}{{{\rm h}^{-1}\msun}}
\newcommand{\hmpc}{{{\rm h}^{-1}\, {\rm Mpc}}}
\newcommand{\dd}{{\rm d}}

\begin{document}
\slugcomment{{\em submitted to Astrophysical Journal Letters}}

\shorttitle{The Epoch of Reionization}

\shortauthors{SOMERVILLE, BULLOCK \& LIVIO} 

\title{The Epoch of Reionization \\
in Models with Reduced Small Scale Power}

\author{Rachel S. Somerville\altaffilmark{1}, James S. Bullock\altaffilmark{2,3}, \& Mario Livio\altaffilmark{1}} 

\altaffiltext{1}{Space Telescope Science Institute, Baltimore MD 21218; 
somerville@stsci.edu; mlivio@stsci.edu}
\altaffiltext{2}{Harvard-Smithsonian Center for Astrophysics, Cambridge, MA 02138; jbullock@cfa.harvard.edu}
\altaffiltext{3}{Hubble Fellow}

\begin{abstract}

Reducing the power on small scales relative to the `standard' \LCDM\
model alleviates a number of possible discrepancies with observations,
and is favored by the recent analysis of \WMAP\ plus galaxy and
Lyman-$\alpha$ forest data. Here, we investigate the epoch of
reionization in several models normalized to \WMAP\ on large scales,
and with sufficiently reduced power on small scales to solve the halo
concentration and substructure problems. These include a tilted model,
the \WMAP\ running-index model, and a Warm Dark Matter model.  We
assume that the Universe was reionized by stellar sources comprised of
a combination of supermassive ($\sim 200 M_{\odot}$) Pop III stars and
Pop II stars with a `normal' IMF.  We find that in all of these
models, structure formation and hence reionization occurs late,
certainly at redshifts below ten, and more probably at $z\la 6$.  This
is inconsistent (at $2\sigma$) with the determination of $z_{\rm
reion} \simeq 17$ from the \WMAP\ temperature-polarization data and is
only marginally consistent with SDSS quasar observations.  The tension
between the galactic-scale observations, which favor low-power models,
and the early reionization favored by \WMAP\ can only be resolved if
the efficiency of Pop III star formation is dramatically higher than
any current estimate, or if there is an exotic population of ionizing
sources such as mini-quasars.  Otherwise, we may have to live with the
standard \LCDM\ power spectrum, and solve the small-scale problems in
some other way.

\end{abstract}

\subjectheadings{cosmology: theory --- galaxies: evolution ---
intergalactic medium}

\section{Introduction}
\label{sec:intro}

The first-year \WMAP\ satellite data have yielded a remarkable
confirmation of what is coming to be regarded as the `Standard Model'
of cosmology: a Hubble parameter $H_0 \simeq 70 $ km s$^{-1}$
Mpc$^{-1}$, matter and vacuum densities consistent with a flat
geometry $\Omega_m \simeq 1 - \Omega_{\Lambda} \simeq 0.3$, and a
nearly scale-invariant primordial power spectrum, $n_s \simeq 1$, at
least on large scales.  There were some surprises, as we will discuss
below, but in many ways the \WMAP\ results highlight the marked
success of standard \LCDM\ in reconciling diverse observations from
the CMB, to galaxy clusters, supernovae, and gravitational lensing.
This remarkable agreement with data on large scales, however, brings
into sharper focus several nagging discrepancies on small-scales.
Specifically, observed galaxy central densities appear to be
significantly lower than the standard \LCDM\ model predicts
\citep{fp94,moore94,abw,swaters03,mcgaugh03,bosch03} and \LCDM\ may
also produce too much substructure compared with observed numbers of
satellite galaxies in the Local Group \citep{klyp99,moore99}.  A
natural way to relieve these problems, while maintaining the
large-scale success of the model, is to reduce the power on small
scales relative to large, either by introducing a `tilt' in the
primordial power spectrum ($n_s < 1$), as expected in some variants of
inflation \citep{abw,zb02}, or by resorting to Warm Dark Matter (WDM)
\citep{ar01,bod01,abw}.  In these models, collapse happens later, when
the Universe was less dense, and galaxy halos are naturally less
centrally concentrated.
Later collapse also alleviates the dwarf problem
\citep{colin00,knebe02,bz:02,zb03}, and the `angular momentum
catastrophe' --- the ongoing problem with producing disk galaxies with
sufficient specific angular momentum in hydrodynamic simulations
within the \LCDM\ framework \citep{sld01}.  Interestingly, one of the
surprises of the \WMAP\ analysis is that it favors (at $\sim 2
\sigma$) a model in which the spectral index varies strongly as a
function of wavenumber, $\dd n/\dd \ln k \simeq -0.03$
\citep{spergel:03}, in precisely the manner needed to cure many of the
problems on small-scales \citep{zb03}.

The second (and perhaps major) surprise from the \WMAP\ report was the
detection of a large amplitude signal in the TE maps, indicating a
large optical depth to Thomson scattering ($\tau = 0.17 \pm
0.04$). The straightforward interpretation of this result is that the
Universe became reionized at $z_{\rm reion} = 17\pm5$
\citep{kogut:03}, rather earlier than many had expected.
Several workers have recently demonstrated that even within the
standard \LCDM\ framework, models of reionization require rather
extreme assumptions in order to produce enough early star or quasar
formation to reionize the Universe by $z\sim 17$
\citep{wl:03,hh:03,cfw:03,sokasian:03,cen:03}.  Put in the context of
the small-scale crises facing the standard model, and indeed, in view
of the running-index model favored by the \WMAP\ analysis, we are left
with a paradox.  The galactic-scale data favor low-power models
precisely because they produce late structure formation, but the high
optical-depth measurement seems instead to favor early collapse.  In
fact, the late reionization implied by low-power models would more
easily accommodate the simplest interpretation of the Gunn-Peterson
troughs observed in several SDSS quasars at $z\ga6$
\citep{fan:01,becker:01}, and the high temperature of the IGM at
$z\sim4$ \citep{hui:03}.

In this {\it Letter} we set out to \emph{quantify} these concerns
using a simple yet conservative model that is informed and motivated
by numerical simulations.  Our aim is to compute conservative upper
limits on $z_{\rm reion}$ for several models that have small-scale
power reduced to plausibly solve the galactic-scale difficulties,
including the running spectral index model favored by \WMAP. In what
follows, we assume that reionization is due to stellar sources alone,
as AGN are likely to be unimportant to reionization at high redshift
unless there are exotic objects (such as 'mini-quasars') that is
completely disjoint from the known population \citep{sb03}.

\section{Models and Normalization}

In order to focus on the ramifications of reducing the small-scale
power, we will fix the cosmological parameters in all of our models to
canonical values: $\Omega_m = 1.0 - \Omega_{\Lambda} = 0.3$, $\Omega_b
h^2 = 0.02$, and $H_0 = 70 $ km s$^{-1}$ Mpc$^{-1}$.  Additional
parameters of each of the models are described in
Table~\ref{tab:models}.  In all cases, we assume that the primordial
power spectrum takes the form $P(k) \propto k^{n_s}$, with a spectral
index that can vary as function of scale as: $n_s = n_s(k_0) + 0.5\,
\dd n/\dd \ln k\, \ln(k/k_0)$ \citep[e.g.][]{kosowsky:95,hhvh:02}.  In
the second column of Table~\ref{tab:models}, we list $n_s$, evaluated
at $k_0 = k_{WMAP} \equiv 0.05$ Mpc$^{-1}$ and in column 3, we give
$\dd n/\dd \ln k$.  In order to make contact with previous work, our
specific choice for normalizing standard \LCDM\ (with $n_s = 1.0$) is
$\sigma_8 = 0.9$, where $\sigma_8$ (column 4) is the linear, rms
fluctuation amplitude of the power spectrum within spheres of radius
$8 \hmpc$, evaluated at $z=0$.  This spectral index and normalization
are very close to the best-fit parameters derived from the \WMAP\ data
alone, without combination with other data sets.  In order to ensure
consistency on large scales, we have chosen to fix the normalization
of the rest of our models to match that of standard \LCDM\ at
$k=k_{WMAP}$.

The first of our models with reduced small-scale fluctuations,
``TILT'', has $\sigma_8 = 0.75$ and modest tilt, $n_s = 0.95$, as
expected in many simple models of inflation (see, e.g. \citeauthor{kinney03}
\citeyear{kinney03}). 
The specific parameter choice is motivated by the work of
\citeauthor{zb02} (\citeyear{zb02,zb03}) who concluded that this
normalization/tilt is favored by galaxy rotation curve data, and may
also alleviate the dwarf satellite problem without the need for
differential feedback.  Interestingly, this parameter choice is also
nearly identical to the best-fit power-law model in the \WMAP\ joint
analysis with other CMB data, the 2dF Galaxy Redshift Survey, and
Lyman $\alpha$ forest data \citep{spergel:03}.  The next model,
``RSI'', is motivated by the same joint \WMAP\ analysis, but
represents the case in which they have allowed a Running Spectral
Index.  Their best fit has $n_s=0.93$ and $\dd n/\dd\ln k = -0.03$.
Our normalization for RSI, $\sigma_8=0.81$, is slightly below the
quoted \WMAP\ value (by $\sim 3.5\%$) because of the simple way we
have elected to normalize on large scales, but this difference is
small compared to the quoted uncertainty. This RSI case also does
quite well in explaining the small-scale observations \citep{zb03}.

Our final model is computed in the context of a WDM scenario, in which
the primordial power spectrum is scale-invariant, but free-streaming
damps fluctuations below a characteristic scale set by the WDM
particle mass $m_{\rm w}$:
\begin{equation}
R_f = 0.2\, (\Omega_{WDM} {\rm h}^2)^{1/3}\left( \frac{m_{\rm w}}{\rm
keV} \right)^{-4/3} {\rm Mpc}.
\end{equation}
We calculate the WDM spectra assuming the same flat cosmology with
 $\Omega_{\rm w}=\Omega_m = 0.3$, and use the approximate WDM transfer
function given by \citet{bbks:86}:
\begin{equation}
P(k) = \exp[-k R_f - (k R_f)^2] P_{CDM}(k).
\end{equation}
We choose $m_{\rm w} = 1.5$ keV, or $R_f \simeq 0.027$ Mpc, in order
to suppress power below a mass scale of $\sim 10^{10} M_{\odot}$.
This choice of WDM mass (with $n_s = 1.0$)
significantly alleviates the usual small scale problems
\citep{ens,abw,zb03}, yet is not ruled out by Ly-$\alpha$ forest data
\citep{narayanan00} or by pre-\WMAP\ constraints from reionization and
early structure formation \citep{bho:01}. We neglect the finite
particle velocity dispersion, as well as associated phase-space
restrictions, which are expected to have negligible consequence for a
1.5 keV model \citep{abw,zent03}.

\section{Halo Collapse Rates}

\begin{figure}
\plotone{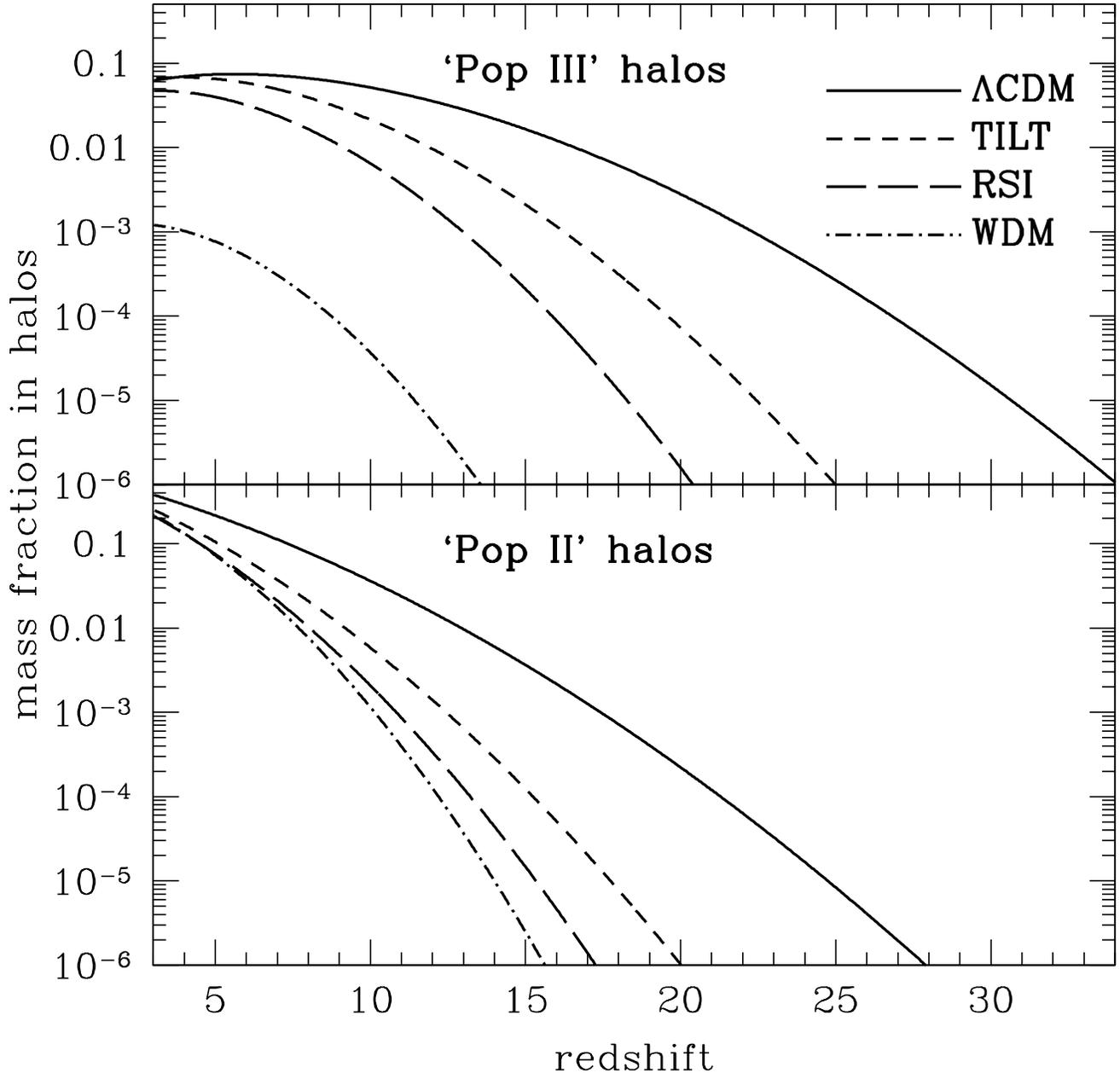}
\caption{\small The fraction of mass in collapsed dark matter halos as
a function of redshift for the four power spectra summarized in
Table~\ref{tab:models}. The top panel shows the mass fraction in halos
with mass greater than $1.0\times10^6\hmsun$ but virial temperature
$T_{\rm vir} < 10^4$ K. In the absence of metals, gas in these halos
can cool primarily by molecular hydrogen, and we associate them with
possible sites of massive ($200 \msun$) Pop III star formation. The
bottom panel shows the mass fraction in halos with $T_{\rm vir}>10^4$
K. We associate these halos with H$_I$ cooling and Pop II star
formation with a `normal' IMF. The impact of reduced small scale power
on early star formation on both of these mass scales is dramatic.
\label{fig:fhalo}}
\end{figure}

We show in Fig.~\ref{fig:fhalo} the fraction of the total mass in the
Universe that is in collapsed halos in two different mass ranges,
computed using the Press-Schechter formalism
\citep{press-schechter}. In the top panel, we show the fraction of
mass in halos with mass greater than $1.0\times 10^6 \hmsun$, but with
temperature less than $10^4$~K (conversions between mass and
temperature are as in \citeauthor{sp} \citeyear{sp}). As discussed
below, we expect gas in these halos (sometimes called ``mini-halos'')
to be able to cool only via H$_2$, and based on the results of
hydrodynamic simulations by \citet{yoshida:03}, they may potentially
harbor massive Pop III stars. In the bottom panel, we show the mass
fraction in halos with $T_{\rm vir} > 10^4$~K, which should be able to
cool via H$_{\rm I}$ and form Pop II stars with a normal Salpeter-like
IMF. We see that all models with reduced small scale power have
dramatically decreased mass fractions relative to Standard \LCDM\ in
halos of the scale expected to be eligible for star formation at high
redshift. The TILT model shows the smallest decrease on all scales,
while the WDM model shows the most dramatic decrease.

\section{Star Formation and Ionizing Photons}

A basic requirement for star formation is the presence of cold, dense
gas.  The two primary coolants in the early Universe, before the
production of heavy elements, are molecular and atomic
hydrogen. Atomic hydrogen is inefficient at temperatures below $\sim
10^4$ K, implying that molecular hydrogen was probably the main
coolant in the first halos to collapse at $z\sim 20$--30, with
temperatures of a few hundred Kelvin. Recent theoretical work suggests
that the first stars to form in these primordial halos may have been
extremely massive, $\sim 100$--600 $\msun$, but that the efficiency
was rather low, with less than $\sim 1$ percent of the available gas
converted to stars or star clusters \citep{abn:00,abn:02,bromm:02}.

The overall efficiency of early star formation is regulated in
mini-halos by a complex interplay between the destruction of H$_2$ by
UV photons and its catalysis by X-rays
\citep{machacek:01,machacek:03,rgs:01,rgs:02b,cen:02}, and in larger
halos by photo-evaporation and supernova feedback
\citep{ciardi:00,squelch,benson:02}. The efficiency of production and
mixing of heavy elements also determines the epoch at which the IGM
becomes sufficiently polluted with metals to allow cooling to lower
temperatures and fragmentation, leading to a shift from the formation
of solely supermassive stars to a more normal Salpeter-like IMF
\citep{bfcl:01}. For a more detailed discussion of this scenario, see
the excellent review by \citet{loeb-barkana}, as well as recent papers
by \citet{cen:02}, \citet{hh:03}, and \citet{wl:02,wl:03}.

We model the global star formation rate density (SFRD) by assuming
that it is proportional to the rate at which gas collapses into halos
in a given mass range:

\begin{equation}
\dot{\rho}_* = e_* \rho_b \, \frac{{\rm d}F_h}{\rm
dt}(M>M_{\rm crit}),
\label{eqn:sfrA}
\end{equation} 

where $\frac{{\rm d}F_h}{\rm dt}(M>M_{\rm crit})$ is the time
derivative of the fraction of the total mass in collapsed halos with
masses greater than $M_{\rm crit}$, obtained from the halo mass
function $\frac{\dd n_h}{\dd M}(M,z)$ given by the Press-Schechter
model, and $\rho_b$ is the mean density of baryons. We adopt $M_{\rm
crit}=M(T_{\rm vir}=10^4 {\rm K})$ and $e^{\rm II}_{*}=0.1$ for `Pop
II' halos, and $e^{\rm III}_{*}=0.002$ and $M^{\rm III}_{\rm crit}=1.0
\times 10^6 \hmsun$ for `Pop III' halos. In \citet[][SL03]{sl:03}, we
found that this simple prescription, with these parameter values,
agrees well in the redshift range $3 \la z \la 30$ with more detailed
semi-analytic models and hydrodynamic simulations of Pop II star
formation including photoionization and supernova feedback
\citep[e.g.][]{spf,sh:03}, and also with the general analytic
arguments outlined by \citet{hs:02}. The global star formation rate
predicted by this approach is also consistent with observational
constraints at $3 \la z \la 6$ (SL03). Similarly, our Pop III SFRD is
in good agreement with the detailed numerical hydrodynamic simulations
of Pop III formation by \citet{yoshida:03}, as also shown explicitly
in SL03. In both cases, the analytic recipe used here slightly
\emph{overestimates} the star formation rate relative to the more
realistic simulations, thus leading to more optimistic results for
early reionization.  As we do not know when the transition from solely
supermassive Pop III stars to Pop II star formation will occur, we
conservatively shut off the Pop III mode at $z<6$ (when the IGM is
known to be significantly polluted with metals). Earlier shut-off
times will only lead to later reionization, making our conclusions
even stronger.

For Pop II stars, we use the results of \citet{leitherer:99} for the
number of $\lambda < 912$ \AA\ photons produced by low metallicity
stars with a ``bottom-light'' Salpeter IMF. At ages less than about 3
million years, these stars produce about $dn_{\gamma}/dt = 8.9 \times
10^{46}$ photons s$^{-1}\, M^{-1}_{\odot}$. For Pop III, we assume
that each star produces $1.6 \times 10^{48}$ photons s$^{-1}
M^{-1}_{\odot}$ for a lifetime of 3 million years \citep{bkl:01}. We
note that, in the spirit of optimism that we have adopted throughout,
these values are at the high end of the estimates of ionizing photon
production and lifetime for both populations.

\begin{figure} 
\plotone{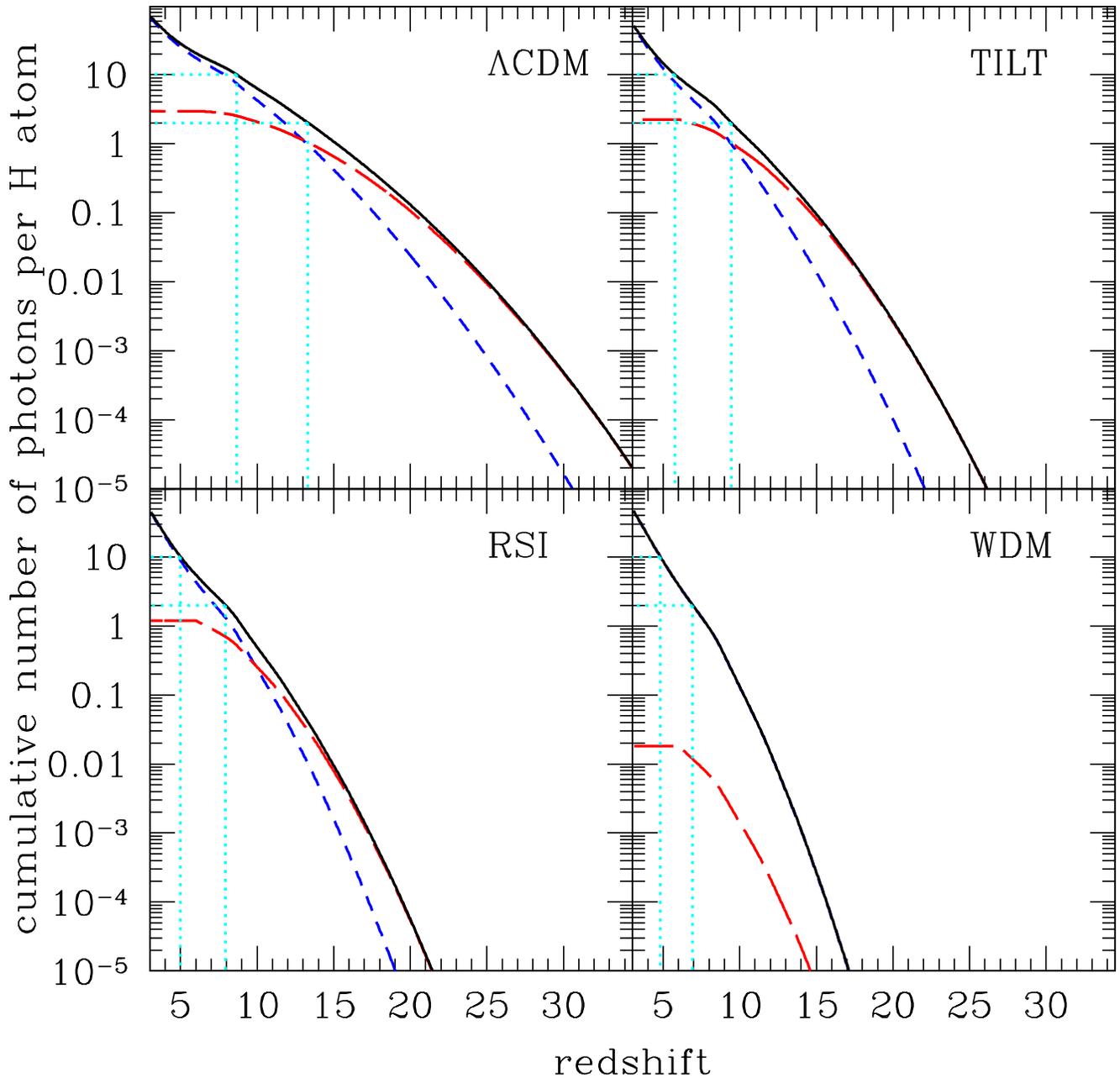}
\caption{\small The cumulative number of hydrogen-ionizing photons per
hydrogen atom produced, as a function of redshift, for the four
initial power spectra. Long dashed lines show the contribution from
Pop III stars, short dashed lines show the contribution from Pop II
stars, and solid lines show the total. Straight dotted lines indicate
the redshift at which 2 and 10 ionizing photons per atom have been
produced, and roughly bracket the range of redshifts when the overlap
of ionized regions is expected to occur. In all three models with
reduced small scale power, overlap occurs later than $z\sim10$. This
may be compared with the estimate of $z_{\rm reion} = 17 \pm 5$ from
the \WMAP\ temperature-polarization results \protect\citep{kogut:03}. 
\label{fig:nphot}}
\end{figure}

The cumulative number of ionizing photons per hydrogen atom in the
Universe is shown in Fig.~\ref{fig:nphot}, for each of the four models
summarized in Table~\ref{tab:models}. The ionized fraction $x_e$ is
expected to scale as this quantity times $f_{\rm esc} \, f_{\rm ion}/
C_{\rm clump}$ \citep[e.g.,][]{spergel:03}, where $f_{\rm ion}$ is the
number of ionizations per UV photon, $f_{\rm esc}$ is the fraction of
ionizing photons that escape from the galaxy, and $C_{\rm clump}$ is
the clumping factor, reflecting the clumpiness of the IGM.
Many analytic and numerical studies have addressed the issue of the
precise number of ionizing photons per atom needed to attain
`overlap', or reionization that is $\sim99$ percent complete
\citep{sokasian:03,csw:03,razoumov:02,haiman:01,miralda-escude:00,gnedin:00b,madau:99}.
This number is a function of the local IGM density (thus, indirectly,
of the redshift of reionization), and of the masses of the halos
harboring the sources, as more massive sources are expected to be in
denser environments with higher clumping factors.
For reference, \citet{sokasian:03} find (in the highest resolution
numerical simulation of reionization by stellar sources to date) that
a gross budget (i.e., before accounting for the escape fraction) of
about 5--20 photons per H atom was required to achieve a
volume-weighted ionization fraction of 99\%. The clumping factors in
numerical simulations are known to be potentially overestimated due to
the limited numerical resolution \citep{haiman:01,sokasian:03}, and
no numerical simulation to date has included the contribution from Pop
III stars, which are likely to be less clustered. We therefore
entertain the possibility that as few as $\sim 2$ ionizing photons per
H atom may be able to do the job.  The redshift at which a total of 2
or 10 ionizing photons per H atom have been produced in each of the
models is shown in Fig.~\ref{fig:nphot}, and recorded in
Table~\ref{tab:models}. We regard these values as bracketing a
generous but plausible range for the expected redshift of overlap in
these models.

One can quickly see that it is difficult to achieve reionization
earlier than $z\sim10$ in any of the models with reduced small scale
power. In the RSI model favored by \citet{spergel:03}, the number of
ionizing photons per hydrogen atom $n_{\gamma}$ reaches the most
optimistic range for expected overlap ($n_{\gamma} \ga 2$) at $z\sim
8$ --- only slightly earlier than the redshift at which the WDM model
reaches this range ($z\sim7$). Values of $n_{\gamma} \ga 10$--20,
corresponding to more typically favored values of the photon escape
fraction ($f_{\rm esc} \sim 0.2$), recombination rate, and clumping
factor, are attained only at $z \sim 5.8$--4.8 in the models with
reduced small scale power --- thus these models (particularly WDM and
RSI) may not even be able to reionize the Universe early enough for
consistency with the \SDSS\ quasar observations.  As noted above, we
have allowed massive Pop III star formation to continue until $z=6$,
although we actually expect it to shut off much earlier. This would
lead to even later reionization, as shown by the short-dashed lines in
Fig.~\ref{fig:nphot}, which show the contribution from Pop II stars
only.

\subsection{Extreme Parameter Values}
Certainly it may be argued that the IMF and formation efficiency of
early generations of stars are highly uncertain. We now consider how
much we would have to vary the free parameters in our simple model to
obtain reionization by $z\sim17$ in the RSI model favored by the
analysis of \WMAP\ combined with other data \citep{spergel:03}. These
results are summarized in Table~\ref{tab:extreme}.  In model RSI.x1,
we increase the efficiency of star formation in `Pop II' halos to
100\% (corresponding to all baryons forming stars the moment they are
within a halo) and the assumed mass of Pop III stars to 600 $\msun$,
the maximum expected value suggested by \citet{omukai:03}, though this
is larger than the maximum value advocated by
\citet{abn:02}. Reionization is then expected between $z\sim
10.5$--8. Only if we assume 100\% star formation efficiency in Pop III
halos as well (model RSI.x2) do we find that reionization could
plausibly occur by $z\sim 16$--14, in reasonable agreement with the
\WMAP\ TE results.

Alternatively, we can leave the star formation efficiency parameters
at their fiducial values and vary the number of ionizing photons
produced by each population. For example, the IMF of early Pop II
stars might be top-heavy, because of the higher temperatures and
pressures and lower metallicities at early times
\citep{larson:98}. Moreover, \citet{tsv} find that zero-metallicity
Pop II stars with a Salpeter IMF produce only about 50 \% more
hydrogen-ionizing photons than the low-metallicity
\citet{leitherer:99} models used here \citep[see also][]{schaerer:03}.
We find that increasing the number of ionizing photons per solar mass
of Pop II stars by a factor of 2--10 can only shift reionization to at
best $z_{\rm reion}\sim 10$ (models RSI.x3 and RSI.x4). Even with the
extreme assumption that Pop II stars produce as many ionizing photons
per unit mass as supermassive Pop III stars (about 20 times the
fiducial value), we find it is unlikely that the Universe could be
reionized earlier than $z\sim 11$ in the context of the \WMAP\ RSI
model.

\section{Discussion and Conclusions}
We have investigated early structure formation in the Standard \LCDM\
model and in three models with reduced small scale power, each of
which alleviates conflicts with observations on sub-galactic scales at
low redshift. All the models we considered are consistent with the
\WMAP\ data on large scales, and two of them (TILT and RSI) are
favored by the combined analysis of \WMAP\ and other CMB data with
Lyman-$\alpha$ forest and 2dFGRS data \citep{spergel:03}. We model
star formation using a simple recipe that has been calibrated against
more detailed semi-analytic and numerical hydrodynamic simulations of
Pop II and massive Pop III star formation
\citep{sl:03,sh:03,yoshida:03}. Using these fiducial parameter values,
and assuming that a gross production of between 2--10 ionizing photons
per H atom is needed to attain overlap, we estimate that the Universe
would become reionized between $z\sim13$ and $z\sim9$ with the
Standard \LCDM\ power spectrum, in agreement with several other
semi-analytic and numerical studies of reionization
\citep{gnedin:00b,razoumov:02,hh:03,cfw:03,sokasian:03}. Using the
same approach, we find that \emph{none} of the models with reduced
small scale power can produce at least two ionizing photons per atom
before $z\sim9.5$, which is about 1.5 $\sigma$ lower than the epoch of
reionization favored by \citet{kogut:03} based on the \WMAP\ TE
measurement. More plausible values of $n_{\gamma} \sim 10$ are not
attained until $z\la 6$ in the models with reduced power on small
scales. The WDM and RSI models may have difficulty reionizing the
Universe even by $z\ga 6$, as required by observations of high
redshift quasars \citep{fan:01}. Of the three models we considered,
the model with a fixed tilt $n_s=0.95$ (TILT) is the best candidate
for obtaining a compromise between the requirements of early structure
formation and observed galaxy central densities and sub-structure at
low redshift. The RSI model favored by \citet{spergel:03} is only
marginally better than WDM in terms of the reionization
constraints. It is also worth noting that any attempts to push
reionization to higher redshift by \emph{increasing} the small scale
power (e.g., by adopting a `red tilt' $n > 1$ as suggested by
\citeauthor{cen:03} \citeyear{cen:03}) would further exacerbate the
problems on sub-galactic scales.

We find that varying the efficiency of star formation or the stellar
IMF within a reasonable range of values cannot significantly change
these conclusions. Only if the efficiency of star formation is pushed
to an extreme upper limit (100$\%$ of baryons in halos with $M>
1\times 10^6 \hmsun$ turn instantly into stars) can the RSI model
plausibly reionize the Universe by $z \sim 16$--14. This result is in
agreement with a similar analysis performed by \citet{hh:03}. If such
extreme star formation were allowed to continue to lower redshift
(even to $z\sim4$), such a model would be in clear conflict with
observations of galaxies and the IGM. Even stronger constraints will
soon be obtained from direct observations of $z>6$ galaxies with HST
and the planned JWST telescope.

We conclude that if we require that the Universe was reionized at
least by $z\sim12$, within $\sim 1\sigma$ of the \WMAP\ result, then,
in the context of the current framework, this paradox can be resolved
only by adopting one or more of the following: (1) the efficiency of
Pop III star formation is \emph{much} higher than current theory
suggests, (2) there is an additional population of ionizing sources
(such as mini-quasars) at high redshift, (3) the power spectrum has
some higher-order feature that produces an upturn in power at masses
just below the scale of dwarf galaxies, or (4) we must retain the
scale-invariant, $n=1$ standard \LCDM\ power spectrum, and somehow
solve the small-scale problems in another way.  While none of these
solutions is particularly attractive, we tend to favor number (4) from
the context of reionization, although this has its own difficulties.
If reionization indeed occurred early, and the small-scale problems
are as robust as they appear, then there are significant gaps in our
theoretical understanding of first light \emph{and} the structure of
galaxies on kpc scales.  The connection between the two certainly
warrants more examination.

\section*{Acknowledgments}
\begin{small}
We acknowledge Lars Hernquist, Joel Primack, Naoki Yoshida, and Andrew
Zentner for useful discussions.  JSB would like to thank Rosemary Wyse
for generous hospitality during  his visit to Johns Hokins University,
where  a  substantial fraction of this    work was completed.   JSB is
provided  for by NASA   through Hubble Fellowship  grant HF-01146.01-A
from the Space Telescope Science  Institute, which is operated by  the
Association of  Universities for  Research in Astronomy, Incorporated,
under NASA contract NAS5-26555.
\end{small}

\bibliographystyle{apj} 
\bibliography{reion}

\clearpage

\begin{table}
\begin{center}
\caption{Summary of models 
\label{tab:models}}
\begin{tabular}{lcccccc}
\tableline
Model & $n(k_{{WMAP}})$ &  $\dd n/\dd \ln k$ & $\sigma_8$ & 
$m_{\nu}$ (keV) & 
$z(n_{\gamma}=2)$ & $z(n_{\gamma}=10)$ \\ 
\tableline
\LCDM & 1.0 & 0.0 & 0.90  & 0.0 & 13.3 & 8.7 \\ 
TILT & 0.95 & 0.0 & 0.75  & 0.0 & 9.4 & 5.8 \\
RSI & 0.93 & -0.03 & 0.81  & 0.0 & 7.9 & 5.0\\
WDM &  1.0 & 0.0 & 0.89  & 1.5 & 6.9 & 4.8\\
\tableline
\end{tabular}
\end{center}
\end{table}

\begin{table}
\begin{center}
\caption{RSI model variants
\label{tab:extreme}}
\begin{tabular}{lcccccc}
\tableline
model & $e^{\rm III}_{*}$ & $e^{\rm III}_{*}$ & $N^{\rm II}_{\gamma}$  & 
$z(n_{\gamma}=1)$ & $z(n_{\gamma}=2)$ & $z(n_{\gamma}=10)$ \\ 
\tableline
RSI.x1 & 1.0 & 0.006 & $\times$ 1 & 11.4 & 10.5 & 8.6\\
RSI.x2 & 1.0 & 1.0 & $\times$ 1 & 16.5 & 15.8 & 13.9\\
RSI.x3 & 0.1 & 0.002 &  $\times$ 2 & 9.6 & 8.7 & 6.0\\ 
RSI.x4 & 0.1 & 0.002 &  $\times$ 10 & 11.14 & 10.3 & 8.4\\ 
RSI.x5 & 0.1 & 0.002 & $N^{\rm III}_{\gamma}$  & 11.7 & 11.0 & 9.1\\ 
\tableline
\end{tabular}
\end{center}
\end{table}

\end{document}